\documentclass[numsec,modern,medium,namedate]{oup-authoring-template}

\onecolumn

\graphicspath{{Fig/}}
\usepackage{natbib}

\theoremstyle{thmstyleone}%
\newtheorem{theorem}{Theorem}
\theoremstyle{thmstyletwo}%
\theoremstyle{thmstylethree}%

\newtheorem{corollary}{Corollary}
\newtheorem{lemma}{Lemma}
\usepackage[doublespacing]{setspace}
\usepackage[fontsize=12pt]{fontsize}
\title[PRx]{Fast Semiparametric Density Regression with Weight-localized Predictive Recursion}

\author[1]{Jonathan Lin}
\author[2]{Surya Tokdar}

\address[1]{\orgdiv{Department of Statistical Science}, \orgname{Duke University}, \orgaddress{\street{214 Old Chem Bldg}, \postcode{27705}, \state{NC}, \country{United States}}, \href{Email:email-id.com}{jonathan.lin@duke.edu}}
\address[2]{\orgdiv{Department of Statistical Science}, \orgname{Duke University}, \orgaddress{\street{214 Old Chem Bldg}, \postcode{27705}, \state{NC}, \country{United States}}, \href{Email:email-id.com}{surya.tokdar@duke.edu}}

\begin{document}

\journaltitle{Fast Semiparametric Density Regression with Weight-localized Predictive Recursion}
\DOI{DOI HERE}
\copyrightyear{XXXX}
\pubyear{XXXX}
\access{Advance Access Publication Date: Day Month Year}
\appnotes{Original article}







\abstract{Predictive recursion (PR) is a fast algorithm for nonparametric estimation of a mixing density, with connections to sequential Bayesian updating under a Dirichlet process prior and rigorous frequentist consistency guarantees. Extending PR to the regression setting---where one seeks to estimate how a mixing density varies with covariates---is nontrivial: dependent Dirichlet process priors, the natural Bayesian generalization, gives no simple recursive updating formula. We introduce PRx, which overcomes this challenge through combining kernel-based weight localization with the recursive scheme of the original PR algorithm. The algorithm scales linearly in sample size and covariate dimension, completing in seconds to minutes where MCMC-based competitors require hours. Exactly as with ordinary PR, the algorithm produces as a byproduct a likelihood score---the PRMLx---whose maximizer is shown to be a consistent estimator for unmixed parameters. In simulations and case studies PRx produces conditional density estimates competitive with established Bayesian procedures at a fraction of the computational cost, and can also be adapted for  a wide range of statistical applications including Bayesian model comparison and covariate-dependent multiple testing.}


\maketitle
\section{Introduction}
Predictive recursion \citep[PR;][]{newton2002nonparametric, newton1998nonparametric} is a fast algorithm for nonparametrically estimating a mixing density $f = dF/d\mu$ when independent observations $y_1,\ldots,y_n$ are available from the kernel mixture density $m(y) = \int \phi(y|\theta)F(d\theta)$. The algorithm has strong connections to sequential posterior updating of the mixing measure $F$ under a Dirichlet process prior \citep{ferguson1973bayesian, escobar1995bayesian}, but executes in a fraction of the time typically needed for a full Bayesian estimation of $F$ \citep{tokdar2009}. Additionally, in semiparametric applications where the mixing measure $F$ is viewed as a ``nuisance'' parameter to be marginalized out with respect to a Dirichlet process prior, the resulting marginal likelihood is well approximated by the predictive recursion marginal likelihood \citep[PRML;][]{martin2011semiparametric} which can be calculated as a byproduct of PR. These properties have made PR and PRML attractive choices for a variety of semiparametric approximate--Bayesian or empirical Bayesian applications, e.g. large-scale multiple testing \citep{martin2012nonparametric, scott2015false}, post-selection inference \citep{woody2022optimal}, and anytime valid inference \citep{dixit2025anytime}. PR is also of considerable independent interest to the predictive Bayes literature \citep{Fortini2020quasi, fong2023martingale}.

When data comes in the form of outcome variables $y_i$ paired with predictors $x_i$, one might be interested in estimating the conditional density $m(y|x)$ under a mixture formulation: $m(y|x) = \int \phi_\sigma(y|\theta)F(d\theta|x)$, where $\sigma$ collects additional parameters of interest. \cite{maceachern1999dependent, maceachern2000dependent} introduced a broad class of dependent Dirichlet process (DDP) priors for Bayesian modeling of such covariate dependent mixing measure $F(\cdot|x)$. Unlike the ordinary Dirichlet process prior, typical DDP priors consist of a large number of latent variables \citep[see][for a review]{quintana2022dependent}, resulting in a manyfold increase in computational cost and requiring delicate assumptions for good statistical behavior \citep{pati2013posterior}. Arguably, the most practical implementation of DDP is given by probit or logit stick-breaking priors \citep{rodriguez2011nonparametric, ren2011logistic, rigon2021tractable}, but even these implementations rely on MCMC procedures that are prohibitively slow for large data sets \citep{ascolani2025fast} and are sensitive to initialization \citep{hastie2015sampling}. While some alternative conditional density models, relying on Gaussian processes \citep{tokdar2010bayesian} or additive trees \citep{li2023adaptive}, offer prior formulation with fewer latent variables and more tractable theoretical properties, their computational cost exceeds that of the best DDP implementations.

Unlike the case of a simple Dirichlet process prior, no simple formula can be produced for sequential posterior updating under a DDP prior, thus presenting a theoretical bottleneck in extending PR to the covariate-dependent case. In this work we introduce a simple extension of PR that overcomes this challenge with provable asymptotic guarantees and competitive empirical performance (Sections \ref{sec:corp}, \ref{sec:theory}). The principal idea is that of a kernel-based ``weight localization.'' 

The regular PR uses a sequence of decreasing weights which control the speed at which new information presented by observation $y_i$ is utilized in the $i$-th step of the recursion. In our extension, which we call PRx (Section \ref{sec:methodology}), these weights are taken to be covariate-dependent so that the speed of information accumulation takes into account the information presented by the paired $x_i$ observation: the update of the current estimate of $f(\theta|x)$ is influenced more heavily by data pairs $(x_i,y_i)$ for which $x_i$ is near $x$. For a single target $x$ for which we aim to recover $f(\theta|x)$, the algorithm scales as $O(n)$ where $n$ is the sample size of the recorded data. We show that the PRx estimate of $m(y|x)$ is consistent in Kullback-Leibler divergence to the projection of the true conditional density within the mixture model class determined by the kernels $\phi_\sigma$ and the dominating measure $\mu$ (Section \ref{sec:theory}). Furthermore, exactly as with ordinary PR, one can marginalize out the mixing density $f(\theta|x)$ to obtain a likelihood score in the additional model parameters $\sigma$. This score, which we call the PRMLx, is obtained as a byproduct of the main recursion, and for large $n$ approximates the likelihood of the data conditional on the covariates. Therefore, the value $\hat{\sigma}_n$ that maximizes the PRMLx could be seen as a natural estimate of $\sigma$. We provide empirical evidence that this estimate leads to excellent overall performance in choosing an appropriate kernel bandwidth in the weight-localization step of PRx. 

We present three case studies to illustrate the practical relevance and scope of PRx. The first is done on a corporate voting dataset, in which we use PRx to obtain density estimates of voting proportion conditional on some covariates of interest. PRx achieves competitive cross-validated scores when compared to other Bayesian procedures while maintaining a superior runtime. In the second case study we measure the negative impact of smoking on the birth weight of a child by modeling the birth weight as a conditional skew-normal mixture, where the skewness is encoded as a linear function of the smoking indicator. We propose using the PRMLx to estimate the intercept and coefficient in this linear function, as well as to construct approximate Bayes factors for comparison against a reference model. The third case study uses PRx to perform  covariate-dependent multiple testing, which is made possible by the fact that the dominating measure $\mu$ may be freely specified. In particular, one can write a variation of Efron's two-groups model \citep{efron2008microarrays, ignatiadis2021covariate} as a Gaussian mixture dominated by a spike-and-slab-type measure \citep{martin2012nonparametric}. With these case studies we aim to highlight that PRx, in addition to offering a form of fast conditional mixing density estimation, may also be leveraged for other statistical applications of interest.

Taken together, the results of this paper demonstrate that PRx is not only a faster substitute for existing density regression procedures, but a versatile tool whose reach extends far beyond data visualization. The mixture model at the core of PRx is expressive enough to encode domain knowledge through the kernel $\phi_\sigma$ while also achieving consistency and robustness properties via the Kullback-Leibler projection. Furthermore, the PRx algorithm naturally gives rise to powerful inferential tools --- most notably the PRMLx --- whose scope extends far beyond density estimation (see Sections \ref{sec:skew}, \ref{sec:testing}). Where state-of-the-art Bayesian competitors are encumbered by MCMC procedures that scale poorly with sample size and dimension, PRx completes in seconds to minutes while delivering competitive results. This combination of speed, theoretical grounding, and versatility should place PRx as a leading tool when employing semiparametric mixture models. 
\section{The Predictive Recursion Algorithm}
\label{sec:PRalgo}

\noindent
The predictive recursion algorithm described in \cite{newton1998nonparametric} and  \cite{newton2002nonparametric} proceeds as follows: given a sequence of data $y_1,\ldots, y_n$, generated i.i.d. from the mixture distribution $m(y) = \int\phi_\sigma(y|\theta)f(\theta)\mu(d\theta)$ one can proceed in estimating $f(\theta)$ through specifying a sequence of weights $w_i$ and an initial guess $f_0$. To be clear, we index the kernel $\phi_\sigma$ by $\sigma$, which collects all unmixed parameters. The predictive recursion algorithm then iterates forward as:
\begin{equation}
    \begin{split}
        m_{i-1}(y) &= \int \phi_\sigma(y|\theta)f_{i-1} (\theta)\mu(d\theta)\\ \label{eq:pralgo}
        f_i(\theta)&= (1-w_i)f_{i-1}(\theta) + w_i\frac{\phi_\sigma(y_i|\theta)f_{i-1}(\theta)}{m_{i-1}(y)}
    \end{split}
\end{equation}
One can view equation \ref{eq:pralgo} as updating $f_i$ as a mixture between the previous iterate $f_{i-1}$ and a posterior distribution with likelihood $\phi_\sigma$ and prior density $f_{i-1}$. A more informative view is given by Newton's original paper, in which the predictive recursion algorithm is justified as a repeated one-observation Baye's-optimal estimate under a sequence of nested Dirichlet process priors \citep{newton2002nonparametric}. In particular, let us assume that we have access only to one observation, $y_1$, and we posit the model: $y|\theta\sim \phi_\sigma(y|\theta), \hspace{5pt}\theta|G\sim G(\theta),\hspace{5pt}G\sim \text{DP}(\alpha, F_0)$, where $y$ is conditionally i.i.d. given $\theta$ and $\theta$ is conditionally i.i.d. given $G$. Then the posterior distribution of $\theta_2$ given $y_1$ can be written as $f_1$ in equation \ref{eq:pralgo}:
\begin{equation}
    \begin{split}
        \mathbb{P}(\theta_2\in B|y_1)
        &= \underbrace{\frac{\alpha}{\alpha+1}F_0(B) + \frac{1}{\alpha+1}\int_B\frac{\phi_\sigma(y|\theta')f_0(\theta')}{\int \phi_\sigma(y|\theta)f_0(\theta) \mu(d\theta)}\mu(d\theta')}_{F_1}
    \end{split}
\end{equation}
Letting $\alpha = (1-w_1)/w_1$ recovers exactly the distribution function associated with $f_1$ in equation \ref{eq:pralgo}. The algorithm then proceeds by taking $F_1$ to be the base measure of another Dirichlet process prior with rate $\alpha = (1-w_2)/w_2$ and solving for the next Baye's estimate, $\mathbb{P}(\theta_3\in B|y_2)$, acting as if we have only observed $y_2$. This continues for all $i\in[n]$. Notice that, even though we only pretend to observe $y_i$ when calculating the $i$th iterate, $F_i$ is still a function of \textit{all} previous data $y_1,\ldots,y_i$. 

\cite{tokdar2009} show that, under mild regularity conditions, the above procedure is consistent in the sense that $f_n\xrightarrow{w} f$ almost surely. To handle the nuisance parameters $\sigma$, one approach is to maximize $\sigma$ over the pseudo-likelihood function $\prod_{i=1}^{n}m_{i-1}(y_i)$. Notice that this is exactly the likelihood function that would arise if our data had been generated according to the sequence of Dirichlet process priors specified above. That is, an estimate $\hat{\sigma}$ of $\sigma$ is obtained through maximizing the so-called predictive recursion marginal likelihood (PRML) as follows:
\begin{equation}
    \begin{split}
        \hat{\sigma}_n = \operatorname*{arg\,max}_\sigma \prod_{i=1}^n m_{i-1, \sigma}(y_i)
    \end{split}
\end{equation}
This procedure is known to have desirable asymptotic properties \citep{martin2011semiparametric}. In particular, it is shown that if we define $K^*(\sigma) := \inf \{D_{KL}(m, m_{f,\sigma}): f\in \mathbb{F}\}$ where $\mathbb{F}$ is the set of all probability densities that are absolutely continuous with respect to $\mu$, then if $\Theta^* := \{\theta: K^*(\theta) = \inf_\sigma K^*(\sigma)\}$ consists only of a single element, $\sigma_0$, we have that $\hat{\sigma}_n\rightarrow \sigma_0$ almost surely. 

\section{Methodology}
\label{sec:methodology}
\noindent
We now present the methodology behind PRx. Observable data is given by pairs $(x_i, y_i)_{i=1}^n\in (\mathcal{X},\mathcal{Y})^n$, where $x$ is $p$-dimensional, and the kernel $\phi_\sigma$ is known up to a nuisance parameter $\sigma$. For now we take $\sigma$ to be known and use the following model:
\begin{equation}
    \begin{split}
         x_i&\sim  q(\cdot) \text{ i.i.d.}\\ 
         y_i|x&\sim \int \phi_\sigma(\cdot|\theta)f(\theta|x)\mu(d\theta) \text{ i.i.d.}\\
    \end{split}
\end{equation}
The goal of density regression is to recover the conditional density $m(y|x) = \int \phi_\sigma(y|\theta)f(\theta|x)\mu(d\theta)$. This is an easy task given an estimate of the latent mixing density, $\hat{f}(\theta|x)$, and so we focus our attention on obtaining this latter quantity. 

There are many approaches towards localization; a naive approach to estimating $f(\theta|x)$ would be to simply run ordinary predictive recursion on the subsample $\{y_i: \lVert x_i - x\rVert<\epsilon\}$. The resulting estimate is not bad in practice, but lacks smoothness, requires a principled choice of $\epsilon$, and also suffers from the loss in power associated with ignoring the majority of the data. Instead of localizing over subsets of our data, we propose localizing the weights using a smooth kernel. For example, for fixed $b_j$, consider the anisotropic kernel given by $k_b(x,x') = \exp(-\sum_{j=1}^p b_j (x_j - x')^2)$. For any $x\in \mathcal{X}$, let us define $\beta_i(x) = k_b(x_i,x)$. We then redefine our weights; rather than specifying a deterministic sequence $w_i$, we instead make them a function of $x$:
\begin{equation}
    \begin{split}
        v_i(x) &= \beta_i(x)h(\beta_1(x)+\ldots+\beta_i(x)) \\
        h(z)&\asymp z^{-\gamma}\text{ for }\gamma\in(1/2,1]
    \end{split}
\end{equation}
For now we will denote $S_i(x) = \sum_{j=1}^i \beta_j(x)$, and we'll take our weights to be $v_i(x)$. One notable quality of $v_i(x)$ is that it is now a function of $x_i$, and is therefore a random quantity. The algorithm then proceeds analogously to the original predictive recursion procedure: we first calculate $v_i(x)$ for all $i\in[n]$, then posit an initial guess $f_0(\theta|x)$. For most cases in which $\mu$ is the lebesgue measure, one can simply set $f_0$ to be a uniform distribution over its support. After this we iterate forwards according to the recursion given by:
\begin{equation}
    \begin{split}
        f_{i}(\theta|x) = (1-v_i(x))f_{i-1}(\theta|x) + v_i(x) \frac{\phi_\sigma(y_i|\theta)f_{i-1}(\theta|x)}{\int \phi_\sigma(y_i|\theta)f_{i-1}(\theta|x)\mu(d\theta)} \label{eq:prxalgo}
    \end{split}
\end{equation}
One can think of the localization kernel $k_b(x,x')$ as encoding ``how far away'' or ``how relevant'' a point $x$ is from $x'$. The weight $v_i(x)$ then transforms this to a quantity encoding how far away $x_i$ is from $x$, compounded with the relevance of all previous covariates $x_j$, $j<i$ to $x$, with the influence of points that are farther away decaying at the order of $z^{-\gamma}$. In all of our numerical experiments we use $h(z) = (1+z)^{-2/3}$. The nuisance parameter $\sigma$ can be tuned as a preprocessing step by maximizing an extension of the PRML, which we call the PRMLx function: 
\begin{equation}
    \begin{split}
        \hat{\sigma}_n = \operatorname*{arg\,max}_\sigma \prod_{i=1}^n m_{i-1, \sigma}(y_i|x_i)
    \end{split}
\end{equation}
The theoretical validity of using PRx and $\hat{\sigma}_n$ will be discussed more in Section \ref{sec:theory}. The following section presents some numerical studies to compare PRx with other established density regression procedures, as well as some applications.
\section{Scope of PRx: Some Numerical Experiments}
\label{sec:sims}

Here we present several case studies that demonstrate the wide applicability of PRx. Section \ref{sec:corp} focuses on conditional density estimation, while Sections \ref{sec:skew} and \ref{sec:testing} give applications to semiparametric inference tasks --- in this case, skewness regression and covariate-dependent multiple hypothesis testing. Our goal is to emphasize the versatility of PRx as not only a faster variant of conditional density estimation, but as a flexible procedure that may be cleverly adapted to a wide range of statistical problems. 

\subsection{Conditional Density Estimation Performance and Runtime}
\label{sec:corp}
Bayesian density regression generally faces two practical bottlenecks: slow MCMC procedures, and a reliance on initial values that leads to irreproducibility.  \cite{tokdar2025densitydiscontinuityregression} discusses these challenges when trying to fit a logit stick-breaking model to a corporate voting dataset for the purposes of discontinuity detection. With a large sample size, the model leads to noticeably different density estimates between runs, each of which takes a long time to execute. Our goal is to use PRx to obtain faster, more stable results. The dataset we consider records all management-sponsored proposals made by U.S. corporations between the years of $2003$ and $2015$, and furthermore augments these with variables from the Standard \& Poor's Compustat database and the Institutional Brokers' Estimate System database. These additional covariates, collected in a vector \textbf{X}, are grounded in domain knowledge and includes ISS recommendation, firm size, Tobin's Q ratio, analyst coverage, and past stock returns. These are all predictors that are known to affect corporate voting outcomes \citep{aggarwal2023investors, gompers2003corporate, malenko2016role}. These are then paired with an outcome variable $y$, representing the proportion of voters in favor of a proposal. 

\begin{figure}[ht]
\centering
\begin{minipage}{\textwidth}
  \centering
  \includegraphics[width=0.9\textwidth]{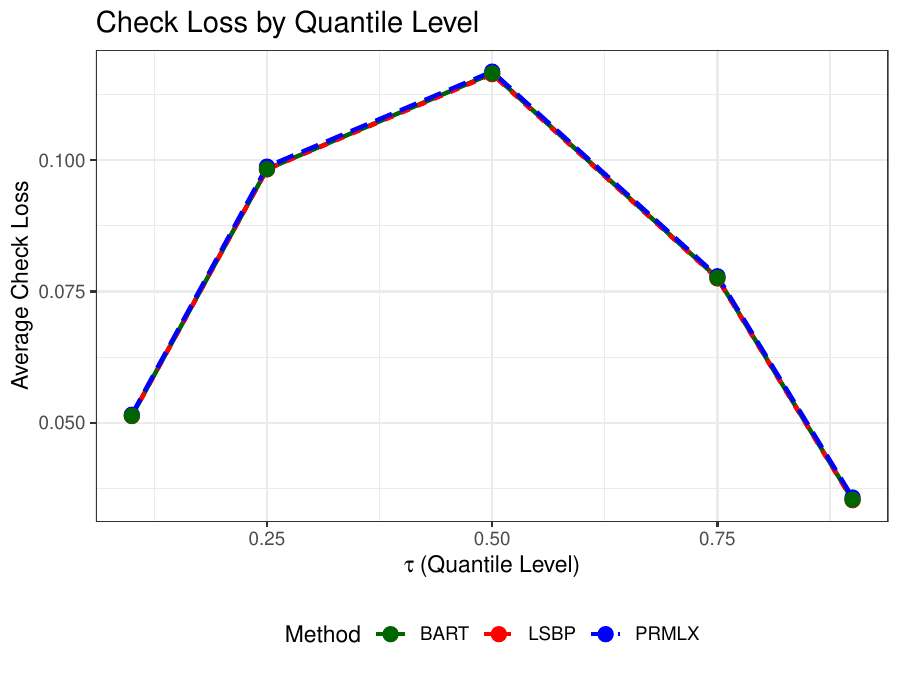}
\end{minipage}
\caption{Cross-validated $\tau$-check scores of PRx, LSBP, and SBART. The three methods result in check scores that are all practically indistinguishable from each other.}
\label{fig:checkscores}
\end{figure}

\begin{figure}[ht]
\centering
{\includegraphics[width=0.45\linewidth]{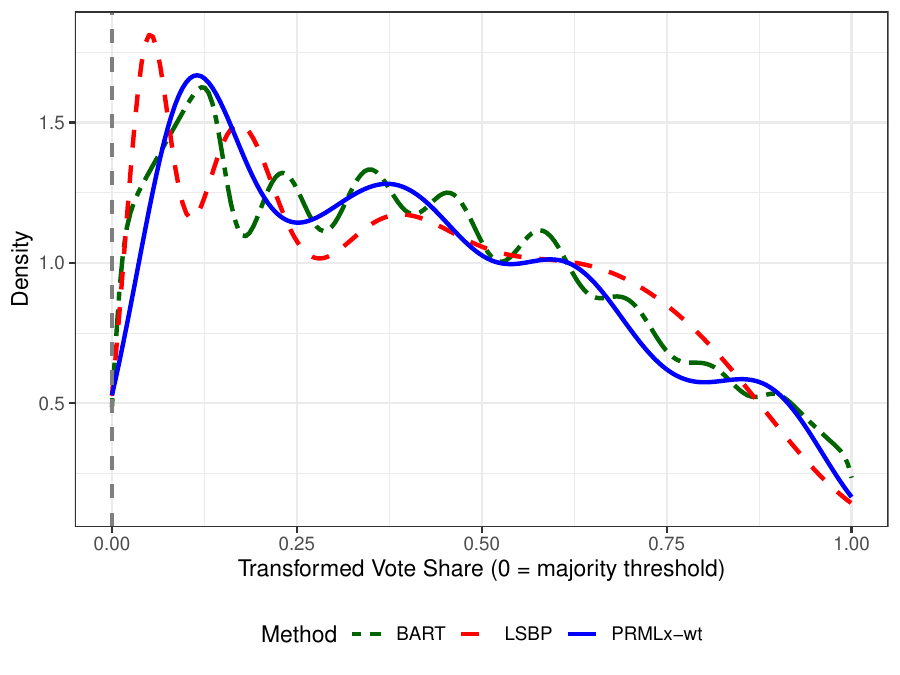}}
\hfil
{\includegraphics[width=0.45\linewidth]{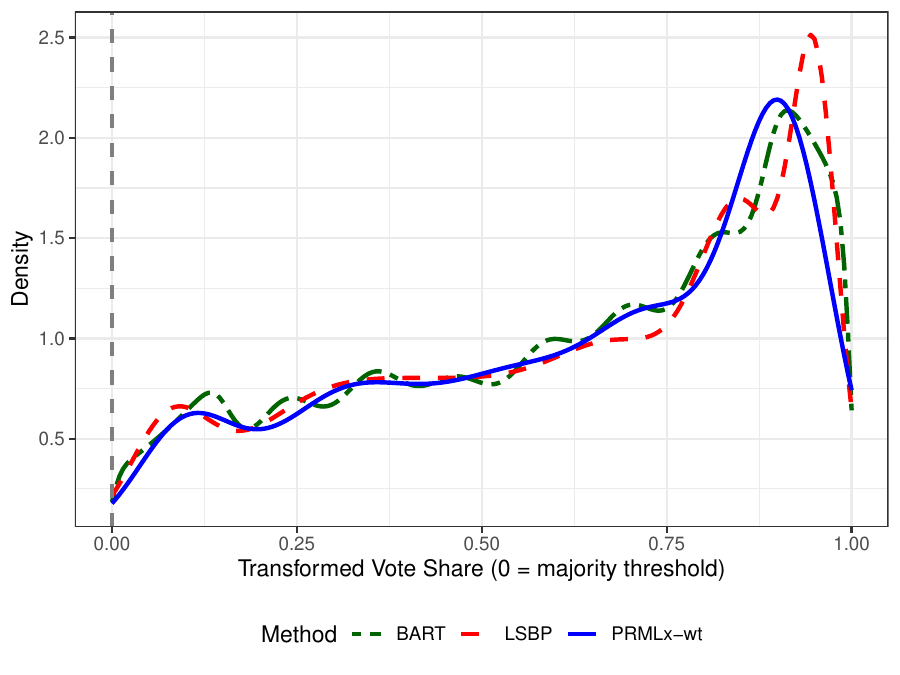}}
{\includegraphics[width=0.45\linewidth]{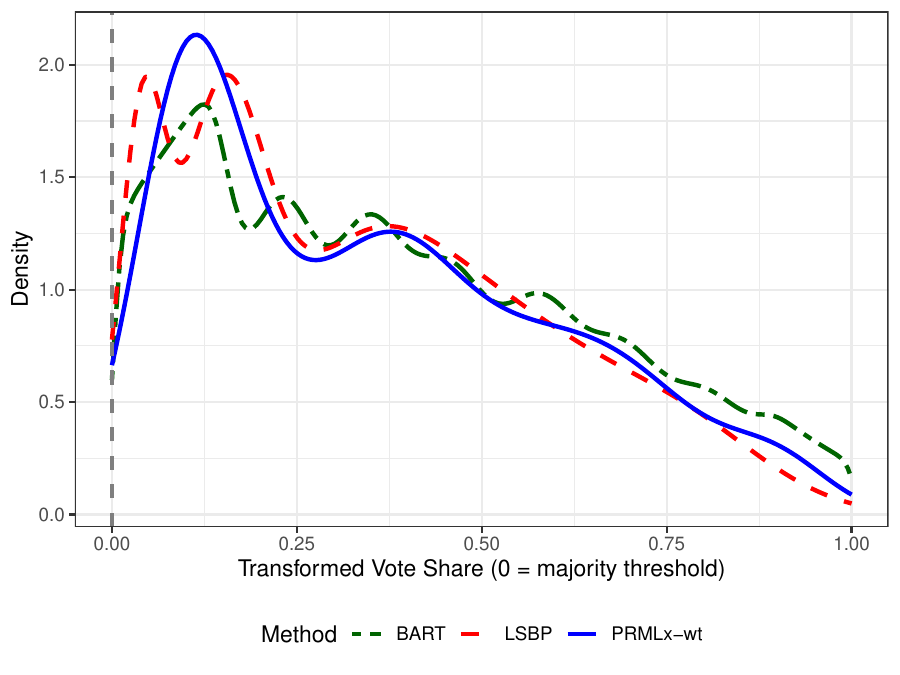}}
\hfil
{\includegraphics[width=0.45\linewidth]{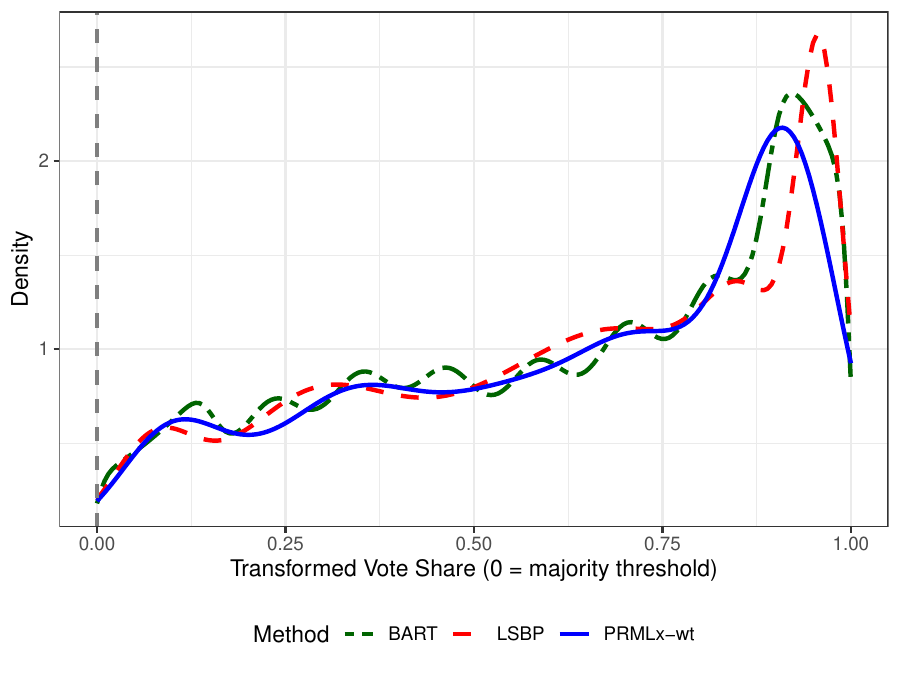}}
\caption{Conditional density estimates at four of the evaluation points. The evaluation points were chosen as the centers of a $k$-means clustering algorithm, with $k=10$.}
\label{fig:corpdensities}
\end{figure}

Following the goals of \cite{tokdar2025densitydiscontinuityregression}, we restrict our attention to all $(\textbf{X}_i, y_i)$ pairs for which $y_i > 0.5$. As a preprocessing step we renormalize these $y$-values to lie on the unit line $[0,1]$. Conditional density estimation is then carried out using PRx, logit stick-breaking \citep{ rigon2021tractable}, and Bayesian additive regression trees \citep{li2023adaptive}. We refer to these latter two methods as LSBP and SBART, respectively. The parametric kernel in PRx is set to be a Gaussian density $\phi_\sigma(y|\theta)$ with mean $\theta$ and variance $\sigma^2$, where $\sigma^2$ is estimated jointly with $\hat{b}_1,\ldots\hat{b}_p$ using PRMLx maximization (Section \ref{sec:theory}) and $f_0$ is initialized as a uniform distribution over $[\min(\textbf{y}) - 1.5\cdot\text{sd}(\textbf{y}), \max(\textbf{y}) + 1.5\cdot\text{sd}(\textbf{y})]$. In order to evaluate goodness of fit, a cross-validation procedure is used in which we split the data into $k=5$ training/test sets. PRx, LSBP, and SBART are used to fit conditional density estimates on the training sets, after which we calculate the \textit{$\tau$-check score} $\text{CS}(\tau)$ at various levels of $\tau$. To be clear this takes the following form:
\begin{equation}
    \begin{split}
        \text{CS}(\tau) = \frac{1}{n_{\text{test}}} \sum_{i \in \text{test}} \rho_\tau(Y_i - \hat{Q}(\tau|X_i))
    \end{split}
\end{equation}

where $\hat{Q}(\tau|\textbf{X}_i)$ is the $\tau$-quantile of the conditional density estimate given by either PRx, LSBP, or SBART, and $\rho_\tau(r) = r\cdot(\tau - 1\{r<0\})$. $\rho_\tau$ is oftentimes referred to as the check loss. The above means of using cross-validation to check goodness of fit was originally used to evaluate the quality of various quantile regression procedures \citep{yang2017joint}. In essence, we are evaluating the quality of our conditional density estimates by how well their quantiles match the observed data. We calculate $\text{CS}(\tau)$ at $\tau = (0.1, 0.25, 0.5, 0.75, 0.9)$ for the three methods considered. We then repeat this process for each of the $k=5$ training/test set pairs. Cross-validation scores can then be calculated by averaging the values of $\text{CS}(\tau)$ over the folds. These are visualized in Figure \ref{fig:checkscores}, and conditional density estimates are given in Figure \ref{fig:corpdensities}. 

After subsetting to all points with $y_i>0.5$, we were left with $n=14505$ data points and a $5$-dimensional covariate vector. The most notable difference between the three methods was the time it took to run the algorithm. While PRx took only $51$ seconds to run predictive recursion over $20$ permutations of the data, LSBP took 20 minutes and SBART took $2.4$ hours to run. Such procedures suffer from slow mixing and difficulty in computing the associated likelihoods, especially in large datasets such as this one. PRx, on the other hand, bypasses these slow computations while maintaining a cross-validated score indistinguishable from those of LSBP and SBART. Furthermore, we found that while LSBP and SBART led to noticeably different results between runs, PRx managed to obtain relatively stable density estimates as a result of permutation-averaging (See Corollary \ref{cor:permutations}).

\begin{figure}[ht]
\centering
\begin{minipage}{\textwidth}
  \centering
  \includegraphics[width=0.9\textwidth]{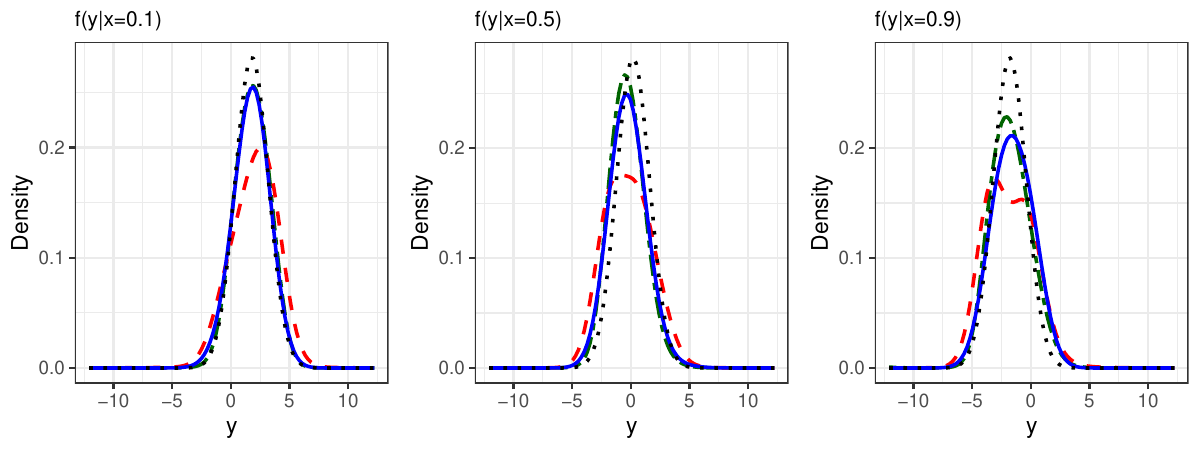}\\
  \includegraphics[width=0.9\textwidth]{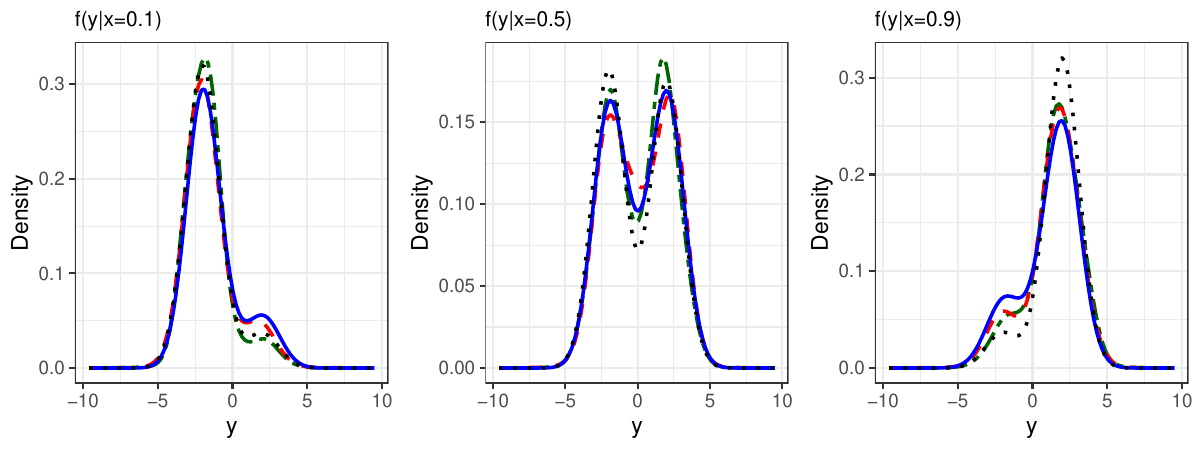}\\
  \includegraphics[width=0.9\textwidth]{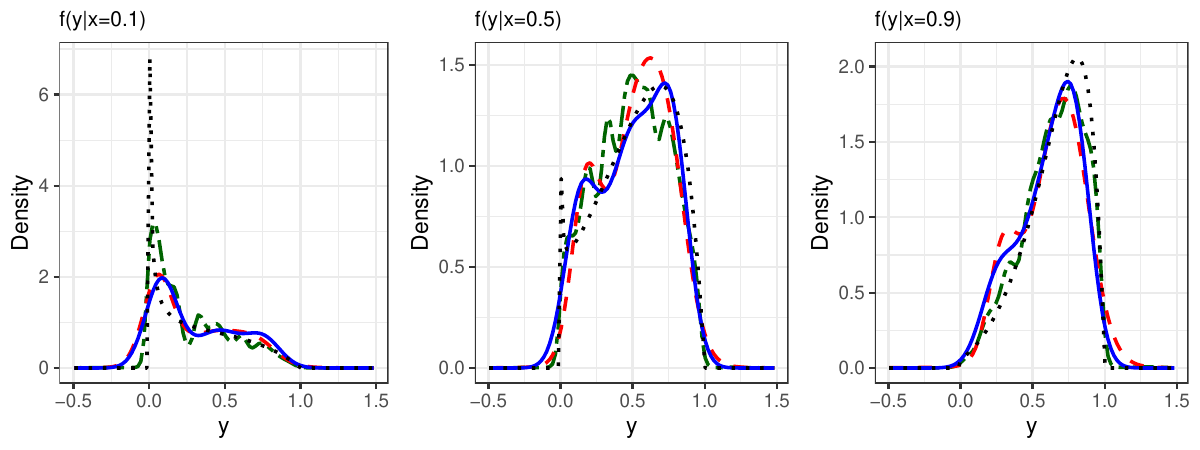}
\end{minipage}
\caption{Conditional density estimates from the first three simulation studies. The first row is taken from the location shift simulation, the second row is taken from the mixture transition simulation, and the third row is taken from the beta concentration simulation. Conditional densities were evaluated at $x = 0.1$, $x = 0.5$, and $x=0.9$. The blue line is the PRx estimate, while the dashed red line is the LSBP estimate and the green dashed line is the SBART estimate.}
\label{fig:simdensities}
\end{figure}

\begin{figure}[ht]
\centering
\begin{minipage}{\textwidth}
  \centering
  \includegraphics[width=0.9\textwidth]{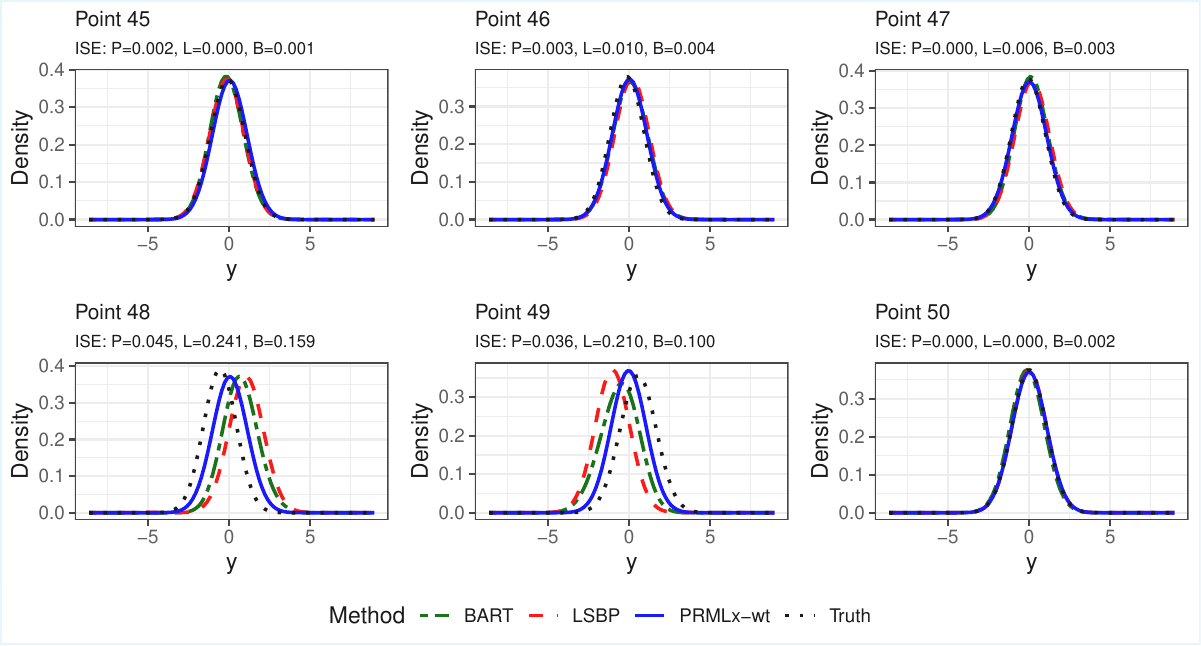}\\
\end{minipage}
\caption{Conditional density estimates from the fourth simulation. The displayed density estimates here are evaluated at three sobol points, the center $(1/2,\ldots,1/2)$, and the corners $(1,\ldots,1)$ and $(0,\ldots,0)$. Points 48 and 49 correspond to corner points in the covariate space.}
\label{fig:simhighdim}
\end{figure}

We now present the application of PRx to four simulations. The first simulation involves data pairs $(x_i,y_i)$ such that $\theta_i|x_i\sim N(3\sin(2\pi x_i), 1)$ and $y_i|\theta_i\sim N(\theta_i,1)$. The second simulation is a mixture-transition model, where we first draw a latent binary variable $c_i|x_i\sim\text{Bernoulli}(x_i)$, and then $\theta_i|c_i=1\sim N(2,1)$ and $\theta_i|c_i=0\sim N(-2,1)$. We then draw $y_i|\theta_i\sim N(\theta_i, 0.5)$. The third simulation involves data in which the kernel is misspecified; $a_i|x_i\sim\text{Gamma}(\alpha(x_i),1)$, where $\alpha(x_i) = 0.5 + 4.5x_i$, and $y_i|a_i\sim \text{Beta}(a_i,2)$. In all three of these settings we let $x_i\sim U[0,1] \text{ i.i.d}$. For the fourth simulation we use a high-dimensional covariate. Let $\textbf{X} = (X_1,\ldots,X_{20})$ be a $20$-dimensional vector where $X_i\sim U[0,1]$ i.i.d. We generate the outcome variable $y_i|\theta_i\sim N(\theta_i, 1)$, where $\theta_i|\textbf{X}_i \sim N(\mu(\textbf{X}_i), \sigma(\textbf{X}_i)^2)$, and:
\begin{equation}
    \begin{split}
        \mu(\textbf{X}) &= \frac{1}{\sqrt{20}}\sum_{i=1}^{20}\bigg[(X_i - 0.5)^3 + 0.3\sin (2\pi X_i)\bigg]\\
\sigma(\textbf{X}) &= 0.2 + \frac{0.3}{20}\sum_{i=1}^{20}X_i
    \end{split}
\end{equation}
We then compare the resulting conditional density estimates from PRx with those obtained by LSBP and SBART. For PRx, we model the conditional density as a conditional mixture over a gaussian kernel: $m(y|x) = \int \phi_\sigma(y|\theta)f(\theta|x)\mu(d\theta)$,
where $\phi_\sigma(y|\theta) = N(\theta, \sigma^2)$. As in the previous subsection, the nuisance parameter $\sigma$ and localization bandwidths $b_1,\ldots,b_p$ are estimated jointly using PRMLx maximization, and   $f_0(\theta|x)$ is initialized as a uniform distribution over $[\min (\textbf{y}) - 1.5 \cdot \text{sd}(\textbf{y}) , \max(\textbf{y}) + 1.5\cdot\text{sd}(\textbf{y})]$. To measure goodness of fit we calculate the mean integrated squared error (MISE) of the conditional density estimates $\hat{m}(y|x)$ over many evaluation points in $\mathcal{X}$. For the first three simulations this takes the form of an evenly-spaced grid over the interval $[0,1]$, while for the high-dimensional simulation we opt to use the corner points $(0,\ldots,0), (1,\ldots, 1)$ followed by a sobol sequence in $20$ dimensions in order to obtain $50$ evaluation points. The sample size of the first three simulations is $n=500$, and the sample size of the high-dimensional simulation is $n=20000$. The results are displayed in Table \ref{tbl:imse_mise}, and conditional density estimates are displayed in Figures \ref{fig:simdensities} and \ref{fig:simhighdim}. 

\begin{table}[ht]
\centering
\begin{tabular}{l l c c}
\hline
Scenario & Method & MISE & Time \\
\hline
Location Shift & PRx & 0.004 & 21s \\
Location Shift & LSBP & 0.016 & 127s \\
Location Shift & SBART & 0.005 & 356s \\
\hline
Mixture Transition & PRx & 0.004 & 28s \\
Mixture Transition & LSBP  & 0.005 & 126s \\
Mixture Transition & SBART  & 0.003 & 169s \\
\hline
Beta Concentration & PRx & 0.211 & 51s \\
Beta Concentration & LSBP & 0.226 & 122s \\
Beta Concentration & SBART & 0.162 & 200s \\
\hline
High Dimensional & PRx & 0.003 & 994s \\
High Dimensional & LSBP & 0.012 & 2360s \\
High Dimensional & SBART & 0.007 & 8392s \\
\end{tabular}
\caption{MISE and runtime across scenarios and methods. PRx outperforms all competing methods in terms of runtime, and results in comparably good estimates.}
\label{tbl:imse_mise}
\end{table}

For all three univariate settings PRx achieves results competitive with those obtained by LSBP and SBART. In the location shift and high dimensional simulations PRx manages to uniformly beat the other two methods in terms of mean integrated squared error. Furthermore, while PRx is beaten by SBART in two of the four simulations, we note that the MISEs are still close in magnitude. In this sense, the estimates given by PRx are comparable to those obtained by other, already-established methods.

The most notable improvement PRx offers is a drastically improved runtime. Where LSBP and SBART required several minutes even for these relatively simple simulations, it took only a few seconds for PRx to run $30$ permutations for exactly the same data. This difference is even more evident in simulation $4$. In particular, PRx required only a few minutes to obtain a conditional density estimate in the high-dimensional setting, whereas LSBP required over half an hour and SBART required a little more than one and a half hours. Indeed, the computing time required to run PRx scales well with sample size, meaning that large datasets do not necessarily lead to prohibitively slow computations, as is commonly seen with MCMC-dependent procedures. As noted in the previous paragraph, this is achieved while maintaining competitive conditional density estimates. 

\subsection{Skewness Regression}
\label{sec:skew}

The model setup of PRx is semiparametric; the user specifies a parametric kernel, which is then convoluted with a nonparametric mixing density. Fully nonparametric density models offer great flexibility, but in many scenarios---especially in the presence of domain knowledge---it is desirable to impose certain structures on the distribution of the outcome variable. For example, it is known by scientists that smoking has a negative impact on the birth weight of a child. In addition to influencing the birth weight mean, one might be interested in determining if smoking negatively skews birth weights. PRx, by virtue of its semiparametric setup, is well-suited to this task; we may select a kernel that explicitly encodes skewness as a parametrized function of a smoking indicator, while permitting the mean to vary flexibly with other covariates through a nonparametric mixing density. The PRMLx may then be used in two ways: we can maximize it to estimate the parameters in the skewness function, and we can construct approximate Bayes factors for comparison against a reference model. The latter allows us to determine whether or not smoking skews birth weights; the former tells us how much smoking skews the birth weights, if it does at all.

To formalize this, let us model the birth weights using a skew-normal mixture. Let $\textbf{X}$ be a multidimensional vector of covariates, and let $t$ be the smoking indicator, which is included in $\textbf{X}$. Let $\phi_{\sigma(t), \psi}(y|\theta)$ be a skew-normal distribution with location $\theta$, scale $\psi$, and skewness $\sigma(t)$. We model $m(y|\textbf{X}) = \int \phi_{\sigma(t), \psi}(y|\theta)f(\theta|\textbf{X})\mu(d\theta)$, where $\sigma(t) = -(\alpha + \beta t)$. In this setting $f(\theta|\textbf{X})$ acts a nuisance parameter that adjusts the location of the skew-normal distribution to account for all covariates $\textbf{X}$. PRMLx maximization can then be used to jointly estimate $\hat{\alpha}, \hat{\beta}, \hat{b}_1,\ldots, \hat{b}_p$, and $\hat{\psi}$. In this setting the estimates $\hat{\alpha}$ and $\hat{\beta}$ are of inferential value to the practitioner, as it gives a notion towards how much smoking skews the underlying distribution. To determine if it affects the skewness at all, we propose a Bayesian hypothesis testing framework. Consider the null model $m^{\mathcal{N}}(y|\textbf{X}) = \int \zeta_\sigma(y|\theta)f(\theta|\textbf{X})\mu(d\theta)$, in which $\zeta_\sigma(\cdot|\theta)$ is a gaussian kernel with variance $\sigma^2$ and mean $\theta$. Let $m_{i-1}^{\text{skew}}(y|\textbf{X}) = \int \phi_{\sigma(t), \psi}(y|\theta)f_{i-1}^{\text{skew}}(\theta|\textbf{X})\mu(d\theta)$ and $m^{\mathcal{N}}_{i-1}(y|\textbf{X}) = \int \zeta_\sigma(y|\theta)f_{i-1}^{\mathcal{N}}(\theta|\textbf{X})\mu(d\theta)$, where $f_i^{\text{skew}}$ and $f_i^{\mathcal{N}}$ denote the PRx estimates of $f(\theta|\textbf{X})$ under the skew-normal mixture model and gaussian mixture model, respectively. We propose constructing approximate conditional likelihood scores $\mathcal{L}_{\text{skew}} = \prod_{i=1}^n m_{i-1}^{\text{skew}}(y_i|\textbf{X}_i)$ and $\mathcal{L}_{\text{norm}} = \prod_{i=1}^n m_{i-1}^{\mathcal{N}}(y_i|\textbf{X}_i)$. By viewing these as proxies for the true conditional likelihood of the observed data, a Bayes factor may be constructed by calculating the ratio $\mathcal{L}_{\text{skew}}/\mathcal{L}_{\text{norm}}$.

\begin{figure}[ht]
\centering

\begin{minipage}[c]{0.42\textwidth}
\centering
\begin{tabular}{lc}
\hline
\textbf{Parameter} & \textbf{Estimate} \\
\hline
$\hat{\alpha}$ & $-2.2856$ \\
$\hat{\beta}$  & $3.7429$ \\
$\hat{\psi}$   & $0.0723$ \\
\hline
\end{tabular}
\end{minipage}
\hfill
\begin{minipage}[c]{0.57\textwidth}
\centering
\includegraphics[width=\textwidth]{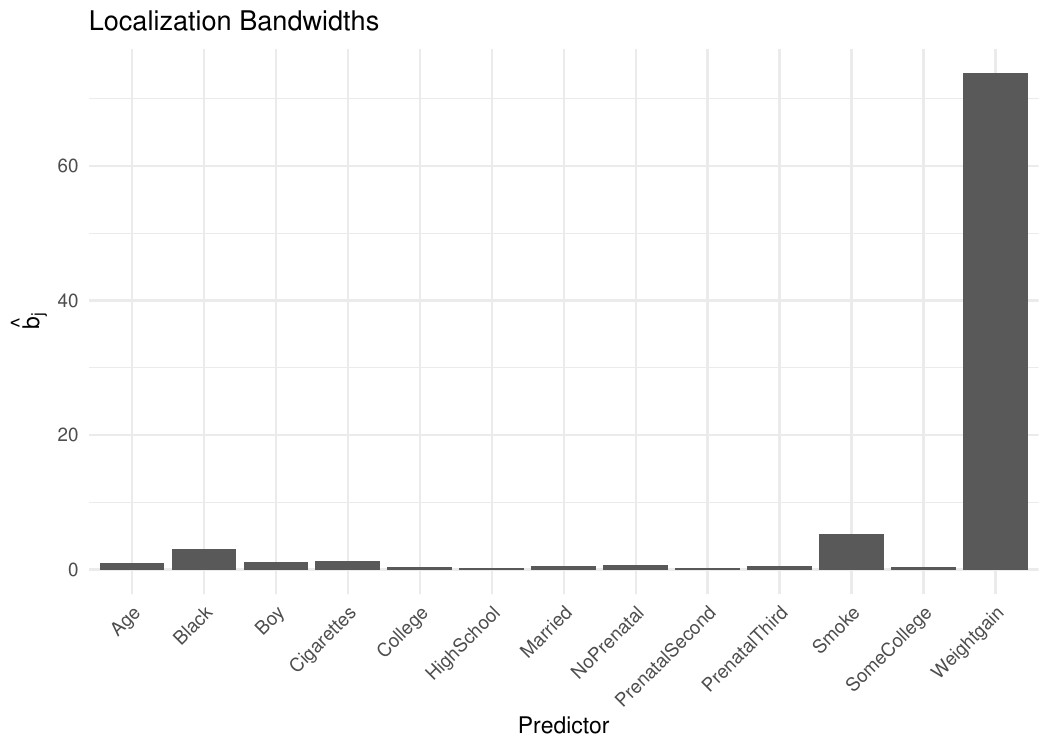}
\end{minipage}

\caption{PRMLx skewness regression parameter estimates (left) and localization bandwidth estimates (right). Notice that the predictors ``Weightgain'', ``smoke'', and ``Black'' have by far the largest estimated values of $\hat{b}_j$, indicating that those three variables require the most localization and thus more heavily influence the mixing density.}
\label{fig:skewreg}
\end{figure}

Using a random subset of $n=15000$ data points from the birth weight data set, we obtain estimates $\hat{\alpha}$, $\hat{\beta}$, $\hat{\psi}$, and $\hat{b}_1,\ldots,\hat{b}_{p}$ for the skew-normal mixture model, and $\hat{\sigma}$ for the gaussian mixture model. As always, the covariate space is normalized to $[0,1]^p$ to make recovery of the localization bandwidths easier. To make the recovery of $\hat{\psi}$ and $\hat{\sigma}$ easier, we also normalize the birth weights $y$ to fit in the unit interval $[0,1]$. The results are presented in Table \ref{fig:skewreg}, and the estimated Bayes factor is calculated to be $\mathcal{L}_{\text{skew}}/\mathcal{L}_{\text{norm}}\approx 4.6\times 10^{7}$, suggesting decisive evidence in favor of using a skew-normal mixture model over the simpler gaussian mixture. This, combined with our estimated value of $\hat{\beta}$, suggests that smoking does indeed skew birth weights towards lower values.
A notable feature of this approach is the complementary roles played by the mixing density and the parametric kernel. Unlike when dealing only with conditional density estimation, the object of interest in this analysis is the kernel; the mixing density serves only as a location adjustment. By imposing certain structures on the kernel, the semiparametric setup of PRx can be leveraged to encode and extract rich dependencies even at the level of the shape of the mixture. In this sense the PRMLx can be considered a powerful inferential tool, as it approximates the likelihood of the observed data. Using this, one can obtain point estimates for parameters that are otherwise difficult to extract, and can construct approximate Bayes factors for Bayesian hypothesis testing and model comparison - all in the presence of a nonparametric mixing density.
\subsection{Multiple Testing with Covariates}
\label{sec:testing}
In most applied settings with continuous outcome variables, it is acceptable to naively set the PRx dominating measure $\mu$ to the lebesgue measure over some compact set. However, when the goal is to encode discrete switching in the distribution of the outcome variable, a more nuanced choice of $\mu$ involving point masses may be useful. We take advantage of this to approach the problem of covariate-dependent multiple testing.

 In this setting, data comes in the form of pairs $(z_i,x_i)$, in which $z_i$ is a $z$-value and $x_i$ is an associated predictor. An extension of Efron's two-groups model \citep{efron2008microarrays, ignatiadis2021covariate} involves a covariate-dependent null probability and a covariate-dependent alternative density. Let $\pi_0(x):\mathcal{X}\rightarrow[0,1]$ denote the covariate-dependent null probability, and let $\gamma_i|x_i\sim \text{Bernoulli}(\pi_0(x_i))$ be a latent variable that encodes whether the $i$th $z$-value comes from the alternative distribution or the null distribution. Then $z_i|\gamma_i=0,x_i\sim f_1(\cdot|x_i)$ and $z_i|\gamma_i=1,x_i\sim f_0(\cdot)$. That is, we assume that the covariate affects the $z$-values at the level of the null probability and the alternative density. Following the work of \cite{efron2005local} and \cite{efron2007size}, we aim to use the covariate-adjusted local false discovery rate $\ell_i = \mathbb{P}(H_i = 0|z_i, x_i)$ in tandem with some thresholding rule in order to control the false discovery rate. By localizing with respect to the covariate in addition to the $z$-value itself (as opposed to only the $z$-value, as in standard methods involving the local false discovery rate), we aim to adjust for the covariate-dependency in both the null proportion and the alternative density when making rejections. The use of $\ell_i$ may also be of independent scientific interest, as it reflects the uncertainty in the hypotheses under specific covariate conditions. 
 
Notice that $\ell_i$ can be expressed as $\ell_i = \pi_0(x_i)f_0(z_i)/f(z_i|x_i)$. Estimation of $\ell_i$ (and controlling the false discovery rate) therefore boils down not only to a problem of conditional density estimation, but also to estimating the function $\pi_0(x)$ in a principled manner. 

We now introduce our model formally. Following \cite{martin2012nonparametric}, we let $z_i|x$ be generated by $f(z|x) = \pi_0(x)\phi(z;\theta,\sigma^2) + (1-\pi_0(x))f_1(z|x)$, and furthermore define $f_1(z|x) = \int \phi(z;\theta+u, \sigma^2)\psi(u|x)du$. $\pi_0(x)$ represents some underlying function of $x$ that maps to $[0,1]$, and $\phi(z;\theta, \sigma^2)$ denotes the density of a gaussian distribution with mean $\theta$ and variance $\sigma^2$. Notice that, under this specification, we can write $f(z|x) = \int \phi(z;\theta+u,\sigma^2)\Psi(du|x)$, where $\Psi(u|x) = \pi_0(x) \delta_0 + (1-\pi_0(x))\psi(u|x)$. Our dominating measure is no longer the lebesgue measure, but is instead given by a point mass at zero with lebesgue measure everywhere else. Such a dominating measure allows us to encode all components of $\ell_i$ within a single mixture model, while retaining Efron's \textit{zero assumption} \citep{efron2008microarrays} by enforcing heavier tails for the alternative distribution. Given an initialized guess, $\Psi_0(u|x)$, one can use PRx to obtain $\hat{\Psi}(u|x)$, an estimate of $\Psi(u|x)$. We can then estimate $\hat{\pi}_0(x) = \hat{\Psi}(0|x)$, and the local false discovery rate at a covariate value $x$ may be calculated as:
\begin{equation}
\begin{split}
\ell(z_i|x) = \frac{\hat{\pi}_0(x) f_0(z_i)}{\hat{f}(z_i|x)}
\end{split}
\end{equation}
In our simulations we assume that the null distribution $f_0$ is a known Gaussian distribution with $\theta = 0$ and $\sigma^2=1$. In applied settings where one feels the need to estimate an empirical null \citep{efron2005local}, PRMLx-maximization may be used to extract $\theta$ and $\sigma^2$ as a preprocessing step. Our rejection rule follows from \cite{sun2007oracle}, in which we first calculate $\ell_i$ for all $(z_i,x_i)$ pairs, order them $\ell_{(1)},\ldots,\ell_{(n)}$, and then calculate $k^*:= \max\{k:\sum_{i=1}^k \ell_{(i)}/k < \alpha\}$. We then reject all hypotheses whose local false discovery rate is beneath $\ell_{(k^*)}$. This rejection rule is known to control the false discovery rate at level $\alpha$ when $\ell_i =\mathbb{P}(H_i=0|z_i)$; we adapt it so that $\ell_i = \mathbb{P}(H_i=0|z_i,x_i)$ instead. While a full theoretical justification of this change is beyond the scope of this paper, we present the following simulation study as evidence of strong empirical performance even in the presence of covariate-dependency.

\begin{figure}[ht]
\centering
\begin{minipage}{\textwidth}
  \centering
  \includegraphics[width=0.9\textwidth]{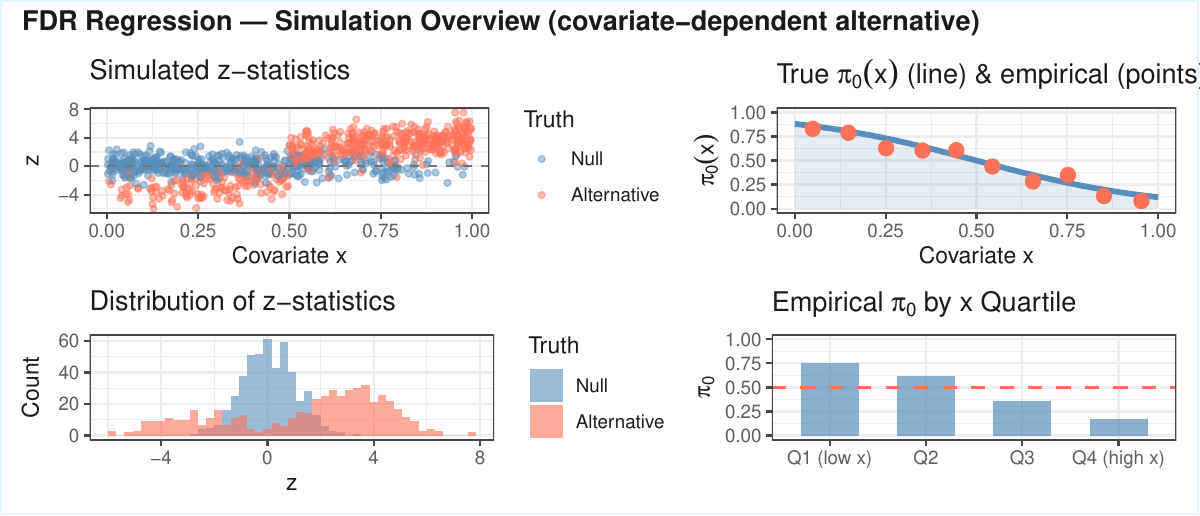}\\
\end{minipage}
\begin{minipage}{\textwidth}
  \centering
  \includegraphics[width=0.9\textwidth]{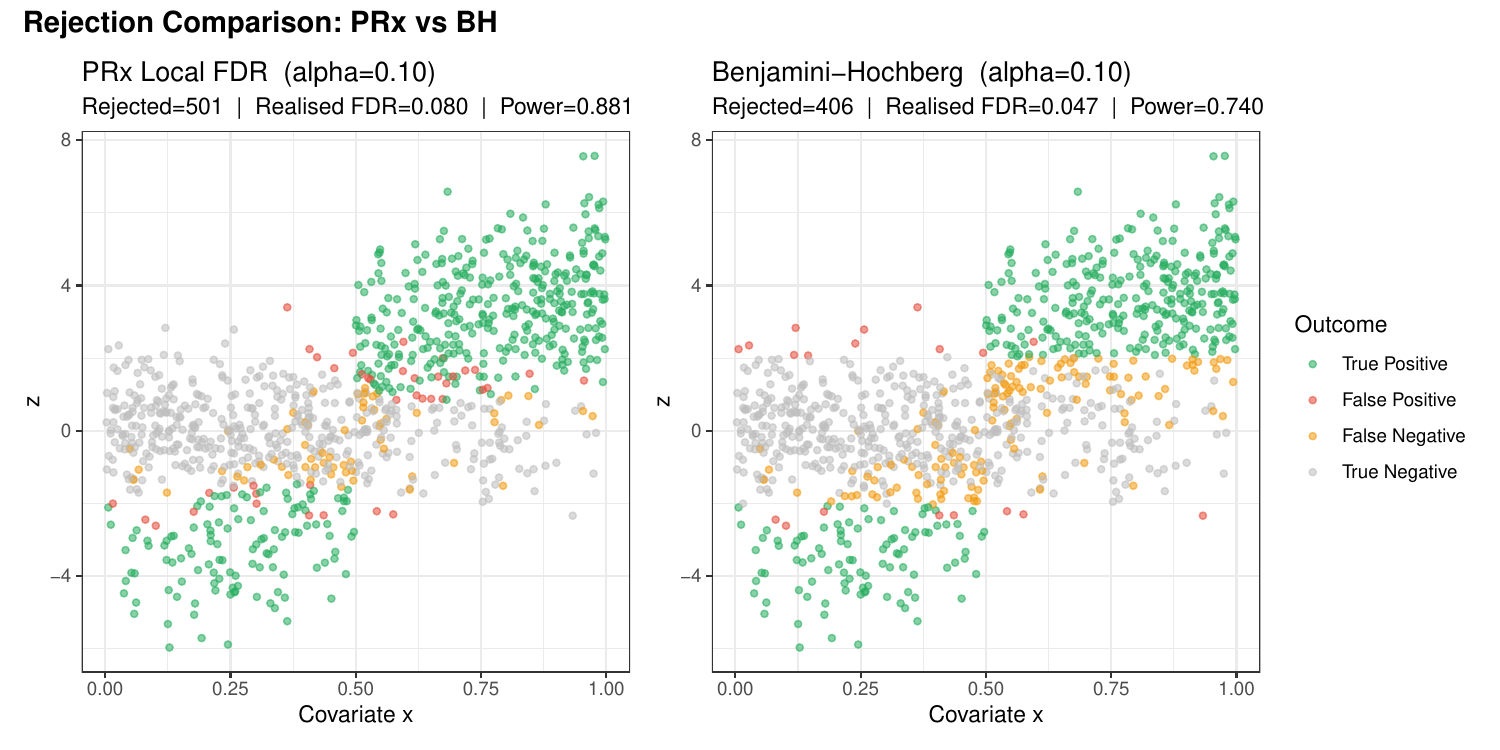}\\
\end{minipage}
\caption{(Top) displays the simulated $z$-values for a single replicate dataset, alongside the true curve $\pi_0(x)$. (Bottom) displays the rejected $z$-values under both PRx and Benjamini-Hochberg for a single replicate dataset.}
\label{fig:testing}
\end{figure}

 We simulate $30$ replicate datasets. In each dataset, $n=1000$ $z$-values are simulated according to the null density $f_0(z) = \phi(z;0,1)$ and alternative density $f_1(z) = \int \phi(z;u, 1)\psi(u|x)\mu(du)$, where $\psi(u|x) = \phi(u; \mu(x), 1)$. The null proportion is given by $\pi_0(x) = 1/(1 + e^{-(2-4x)})$. For this simulation we take $\mu(x) = -4+4x$ for $x\in[0,1/2)$ and $\mu(x)=4x$ for $x\in[1/2,1]$. As always, $\hat{b}$ is estimated using PRMLx maximization. In each replicate, $\Psi_0$ is initialized as $\Psi_0(u|x) = \pi_0 \delta_0 + (1-\pi_0) f_0(u|x)$, where $f_0$ is a uniform distribution on $[-8,8]$ and $\pi_0$ is initialized as $0.75$. The covariate $x_i$ is generated from a uniform $x_i\sim\text{Uniform}[0,1]$ and  our target FDR control level is $\alpha = 0.1$. Figure \ref{fig:testing} visualizes the simulated $z$-values in one such replicate, alongside the rejections made using PRx. To evaluate the results we calculate the empirical false discovery rate by averaging the false discovery proportions between the replicates. These results are compared to those obtained by naively applying the Benjamini-Hochberg procedure \citep{benjamini1995controlling} to the replicate datasets without considering the covariates. 
The simulations give favorable results for PRx. While Benjamini-Hochberg controls the empirical false discovery rate at $0.049$, it attains an average power of $0.737$. PRx, on the other hand, controls the empirical false discovery rate at $0.096$ and attains an average power of $0.888$. The bottom panels of Figure \ref{fig:testing} demonstrate why ignoring the covariates leads to worse performance; while Benjamini-Hochberg is guaranteed to control the false discovery rate, it induces an oversimplified decision rule that rejects only when $|z_i|$ exceeds a fixed threshold. This leads to two crucial shortcomings: first, since the threshold is fixed, rejections are made according to the same rule regardless of the covariate value, whereas the structure of $\pi_0(x)$ informs us that we should be rejecting more aggressively for larger values of $x$. Secondly, it fails to capture the structure of $f_1(z|x)$, which is mean-shifted towards negative values for small $x$ and positive values for large $x$---indeed, the few false discoveries under the Benjamini-Hochberg procedure can largely be attributed to rejections that occur for positive $z$ when $x<0.5$ and negative $z$ when $x>0.5$. Instead of thresholding the $z$-values, PRx thresholds $\ell_i$, which represents the probability of the null hypothesis conditional on both the $z$-values and the covariate. It therefore offers a rejection rule that adapts to the covariate-dependency in both the null proportion and the alternative density. This allows for more aggressive rejections in covariate regimes where the null proportion is small, and adjusts the location of the rejections to be consistent with $f_1(z|x)$.

\section{Consistency Results}
\label{sec:theory}

We have seen in Section \ref{sec:sims} how PRx offers a fast alternative to traditional Bayesian density regression procedures. What remains to be seen is whether or not PRx is consistent - that is, whether or not the PRx estimate $f_n(\theta|x)$ converges to the true latent conditional mixing density $f(\theta|x)$ in some sense. We begin by introducing some assumptions. These are all analogous to the assumptions presented in \cite{tokdar2009}, with the exception of the last one. 
\begin{enumerate}
    \item The sequence $x_1,\ldots,x_n$ are i.i.d. and integrable. Furthermore, $\mathcal{X}$ and $\Theta$ are both compact.
    \item The mapping $\theta\rightarrow\phi_\sigma(y|\theta)$ is bounded and continuous in $\theta$ for each fixed $y$.
    \item The mapping $f(\theta|x)\rightarrow\int\phi_\sigma(y|\theta)f(\theta|x)\mu(d\theta)$ is injective.
    \item For any $\epsilon>0$ and compact set $\mathcal{C}_0\subset\mathcal{Y}$, there exists a compact set $\Theta_0\subset\Theta$ such that $\int_{\mathcal{C}_0}\phi_\sigma(y|\theta)\nu(dy)<\epsilon$ for all $\theta\notin\Theta_0$.
    \item There exists a constant $0<B<\infty$ such that for any $\theta_1,\theta_2,\theta_3\in\Theta$:
    $$\int_\mathcal{Y}\frac{\phi_\sigma(y|\theta_1)^2}{\phi_\sigma(y|\theta_2)^2}\phi_\sigma(y|\theta_3)\nu(dy)<B$$
    \item There exists a constant $C>0$ such that for all $(x_i,y_i)$, it is the case that $m(y_i|x_i)\geq Cm(y_i|x)$. 
\end{enumerate}
We will begin by considering the sums $\sum_i v_i(x)$ and $\sum_i v_i(x)^2$. Recall that in the original predictive recursion algorithm, the deterministic weights $w_i$ are specified so that $\sum_i w_i = \infty$ and $\sum_i w_i^2<\infty$. It turns out that this same fact is also true for the random weights $v_i(x)$. In particular, the following can be shown under assumption 1.
\begin{lemma}\label{lemm:sum}
Under the assumptions laid out above, the weights $v_i(x) = \beta_i(x)h(S_i(x))$ almost surely satisfy:
$$\sum_{i=1}^\infty v_i(x) = \infty\text{ a.s.}\hspace{10pt} \sum_{i=1}^\infty v_i(x)^2<\infty\text{ a.s.}$$
\end{lemma}
While only a small part of the proof of the main theorem, Lemma 1 offers a certain level of comfort in that we are now confident that our weights will satisfy the same conditions as the deterministic weights of ordinary predictive recursion. The main result is now presented:
\begin{theorem}\label{thm:5.2}
    Under the assumptions laid out above, and with weights $v_i(x) = \beta_i(x)h(S_i(x))$, the density estimates $f_n$ obtained from the recursion in \ref{eq:prxalgo} satisfies:
    $$q(x)f_n(\theta|x)\xrightarrow{w} q(x)f(\theta|x)\text{ a.s.}$$
\end{theorem}
The proof of Theorem \ref{thm:5.2} involves decomposing the KL divergence between $m(y|x)$ and $m_n(y|x)$ into a telescoping sum. We provide a cursory outline of the proof here; for more details see the supplementary. First, let $m_n(y|x) = \int \phi_\sigma(y|\theta)f_n(\theta|x)$ and $m(y|x) = \int \phi_\sigma(y|\theta)f(\theta|x)\mu(d\theta)$. We'll denote the $i$-step KL divergence and the $i$-step alternative conditional as:
\begin{equation}
    \begin{split}
        \mathcal{K}_i^*(x) &:= \int m(y|x)\log\bigg(\frac{m(y|x)}{m_i(y|x)}\bigg)\nu(dy)\\
        h_i(y,x)&:=\int\phi_\sigma(y|\theta)\frac{\phi_\sigma(y_i|\theta)f_{i-1}(\theta|x)}{m_{i-1}(y_i|x)}\mu(d\theta)
    \end{split}
\end{equation}
Notice that the conditional density $m_i(y|x)$ now satisfies $m_i(y|x) = (1-v_i(x))m_{i-1}(y|x) + v_i(x) h_i(y,x)$. With this it is possible to express $\mathcal{K}_n^*(x) - \mathcal{K}_0^*(x)$ as a triple sum:
\begin{equation}
    \begin{split}
        \mathcal{K}_n^*(x) - \mathcal{K}_0^*(x) = \underbrace{\sum_{i=1}^n v_i(x) V_i^*(x)}_{(A)} - \underbrace{\sum_{i=1}^nv_i(x) M_i^*(x)}_{(B)} + \underbrace{\sum_{i=1}^nE_i^*(x)}_{(C)}\\ \label{eq:3seriesdecomp}
    \end{split}
\end{equation}
Where 
\begin{equation}
    \begin{split}
        V_i^*(x) &= \bigg(1 - \int m(y|x)\frac{h_i(y,x)}{m_{i-1}(y|x)}\nu(dy)\bigg) + M_i^*(x)\\
        M_i^*(x)&= -\mathbb{E}\bigg[\bigg(1 - \int m(y|x)\frac{h_i(y,x)}{m_{i-1}(y|x)}\nu(dy)\bigg) \bigg| \mathcal{F}_{i-1},x_i\bigg]\\
        E_i^*(x) &= v_i(x)^2 \int p_0(y|x)\bigg(\frac{h_i(y,x)}{m_{i-1}(y|x)}-1\bigg)^2R\bigg(v_i(x)\bigg(\frac{h_i(y,x)}{m_{i-1}(y|x)}-1\bigg)\bigg)\nu(dy)\\
        R(x)&\leq \frac{1}{2}\max\bigg(1, \frac{1}{(1+x)^2}\bigg)
    \end{split}
\end{equation}
A martingale difference argument can be used to show that (A) converges, and the inequality we have with respect to $R(x)$ can be leveraged to argue for the convergence of (C). One can then show that (B) must be positive, and therefore equation \ref{eq:3seriesdecomp} can be rearranged into:
\begin{equation}
    \begin{split}
        \sum_{i=1}^\infty v_i(x)M_i^*(x)&\leq \sum_{i=1}^\infty v_i(x)V_i^*(x) + \sum_{i=1}^\infty E_i^*(x) + \mathcal{K}_0^*(x) < \infty
    \end{split}
\end{equation}
It must therefore be the case that $\mathcal{K}_n^*(x)\rightarrow \mathcal{K}_\infty^*(x)$ for some random variable $\mathcal{K}_\infty^*(x)$. Notice that all of this is done \textit{pointwise} on $x$. A very similar argument can be used for $\mathcal{K}_i(x)$:
\begin{equation}
    \begin{split}
        \mathcal{K}_i(x) = \int f(\theta|x)\log\bigg(\frac{f(\theta|x)}{f_i(\theta|x)}\bigg)\mu(d\theta)
    \end{split}
\end{equation}
One can show that $\mathcal{K}_n(x)\rightarrow \mathcal{K}_\infty(x)$ for some random variable $\mathcal{K}_\infty(x)$. In particular, it is the case that $\sum_{i=1}^\infty v_i(x) M_i(x)<\infty$ almost surely. Additional calculations, however, yield that $M_i(x) \geq C\mathcal{K}_i^*(x)$. Thus, we can conclude that $\sum_{i=1}^\infty v_i(x) \mathcal{K}_i^*(x)\leq 1/C \sum_{i=1}^\infty v_i(x)M_i(x) < \infty$. Since $\sum_{i=1}^\infty v_i(x) =\infty$ almost surely, this implies that $\mathcal{K}_i^*(x)\rightarrow 0$ almost surely on a random subsequence. But we already have that $\mathcal{K}_i^*(x)\rightarrow \mathcal{K}_\infty^*(x)$, and so it follows that $\mathcal{K}_i^*(x)\rightarrow 0$ almost surely. Theorem $3$ from \cite{tokdar2009} then implies weak convergence $f_n(\theta|x)\rightarrow f(\theta|x)$ for fixed $x$. A simple application of the dominated convergence theorem can extend this to weak convergence over the joint space $\mathcal{X}\times \Theta$. 

An issue noted in \cite{tokdar2009} with such consistency results for predictive recursion is that they deal explicitly with large-sample results in an i.i.d. setting, whereas predictive recursion was originally formulated for sequential data, being recursive in nature. The resulting estimate of the mixing density, $f_n$, then depends on the order in which we feed the data into the algorithm. A simple solution exists for this; one could run predictive recursion for many permutations of the observed data and then average the results over the permutations. That is, if $\mathcal{S}$ is the set of all permutations of the indices in $[n]$, then for every $\sigma\in\mathcal{S}$ we run predictive recursion and obtain an estimate of the mixing density, $f_\sigma$. We then obtain our final estimate as $\bar{f}_n := \sum_{\sigma\in\mathcal{S}}f_\sigma / |\mathcal{S}|$. Indeed, it turns out that this strategy is provably consistent as well.
\begin{corollary}\label{cor:permutations}
    $q(x)\bar{f}_n(\theta|x)\xrightarrow{w} q(x)f(\theta|x)$ in probability.
\end{corollary}
Ideally we would run predictive recursion once for every permutation, but in practice we find that only $20$ or $30$ permutations are necessary to achieve stable results. 

Throughout this section we have treated the kernel $\phi_\sigma(y|\theta)$ as a fully known, fixed function. In many applications, however, it is reasonable to expect that either a subset or all of the parameters in $\sigma$ are unknown quantities. Following the predictive recursion maximum likelihood procedure in \cite{martin2011semiparametric}, we propose estimating $\sigma$ by maximizing the PRMLx:
\begin{equation}
    \begin{split}
        L_n(\sigma) = \prod_{i=1}^{n}m_{i-1, \sigma}(y_i|x_i)
    \end{split}
\end{equation}
in order to obtain estimates of $\sigma$. Analogously to the PRML of Section \ref{sec:PRalgo}, this function is exactly the likelihood associated with outcome variables generated from a sequence of Dirichlet process priors, conditional on the covariates. In this sense we can take $L_n(\sigma)$ as a proxy for the true underlying conditional likelihood function, which can act as a score for model comparison and approximate Bayes factors (Section \ref{sec:skew}). One can also show that a normalized version of $L_n$ converges to the expectation of the $\log$-conditional density under the true underlying joint distribution of $(x, y)$:
\begin{theorem}\label{thm:lln}
    $\log L_n/n \rightarrow \mathbb{E}_{(x,y)\sim q(x)m(y|x)}[\log m(y|x)]$ almost surely.
\end{theorem}
This result in and of itself establishes the prediction recursion likelihood function $L_n (\sigma)$ as a quantity of interest, in that it obeys a result analogous to that of the strong law of large numbers. However, Theorem \ref{thm:lln} does not inform us in any way about the behavior of the PRMLx estimate $\hat{\sigma}_n = \text{ arg max}_{\sigma\in \Sigma} L_n(\sigma)$. This is because, in the context of Theorem \ref{thm:lln}, we are assuming that our model formulation is well-specified - that is, allowing $\sigma_0$ to be the true underlying value of $\sigma$, we are assuming that $m (y|x) = \int \phi_{\sigma_0}(y|\theta)f(\theta|x)\mu(d\theta)$, and furthermore, that $m_i(y|x) = \int \phi_{\sigma_0}(y|\theta)f_i(\theta|x)\mu(d\theta)$. In order to analyze the asymptotic behavior of $\hat{\sigma}_n$, we must work in the misspecified setting - that is, we assume that $m_i(y|x) = \int \phi_\sigma(y|\theta)f_i(\theta|x)\mu(d\theta)$ for some value of $\sigma$ that may or may not be correct. 

To be clear, let $\mathbb{F}$ be the space of all measures that are absolutely continuous with respect to $\mu$. Then, for any $F\in \mathbb{F}$, $x\in\mathcal{X}$, and $\sigma\in\Sigma$, we may define $m_{F,\sigma}(y|x) = \int \phi_\sigma(y|\theta)F(\theta|x)\mu(d\theta)$. Firstly, notice that if $f_0\in\mathbb{F}$, then $f_i\in\mathbb{F}$ for all $i = 1,\ldots, n$. Secondly, our choice of $\mathbb{F}$ and $\sigma$ also generates the set $\mathbb{M}_\mathbb{F,\sigma}:=\{m_{F,\sigma}:F\in\mathbb{F}\}$. All of the results up to this point have been derived under the assumption that $m\in\mathbb{M}_\mathbb{F,\sigma}$ - a condition we will now weaken. Instead, we only assume that $m(y|x) = \int\phi_{\sigma_0}(y|\theta)F(\theta|x)\mu(d\theta)$ for some $\sigma_0\in\Sigma$ and $F\in\mathbb{F}$. That is, instead of requiring exact knowledge of $\sigma_0$, we only require input of a compact set in which it lies. Consider the following assumptions:
\begin{enumerate}
    \item $\mathbb{F}$ is pre-compact in the weak topology
    \item $(\theta,\sigma)\rightarrow \phi_\sigma(y|\theta)$ is a bounded and continuous function for almost all $y$.
    \item It is the case that:
    $$\bigg|\frac{\partial }{\partial \sigma}\log m_{i-1}(y|x,\sigma)\bigg|\leq L$$
    Where $L$ is a constant independent of $i$.
    \item The mapping $(\sigma, f)\rightarrow m_{\sigma, f}$ is injective.
    \item There exists a constant $B>0$ such that, for any $\sigma\in\Sigma$:
    $$\sup_{ \theta_1,\theta_2,\theta_3\in\Theta}\int \frac{\phi_{\sigma}(y|\theta_1)^2}{\phi_{\sigma}(y|\theta_2)^2}\phi_{\sigma_0}(y|\theta_3)\nu(dy)<B$$
    \item $\Sigma$ is compact.
\end{enumerate}
Under assumptions (1) and (2), \cite{martin2009asymptotic} show using Kullback-Leibler projections that predictive recursion is robust to model mis-specification. A simple variation of the proof of their main theorem, involving the Robbins-Siegmund almost-supermartingale convergence theorem, shows that $\mathfrak{K}_i^*(x)\rightarrow 0$ almost surely for all fixed $x$, where $\mathfrak{K}^*_i(x) = \mathcal{K}_{i}^*(x) - \mathcal{K}_{\mathbb{F}}$, and $\mathcal{K}_{\mathbb{F}} := \inf \{D_{KL}(m, m_f):f\in\bar{\mathbb{F}}\}$. $\bar{\mathbb{F}}$ is the closure of the set $\mathbb{F}$ in the weak topology. This, combined with assumptions (3)-(5), allows us to establish the consistency of the PRMLx estimate.
\begin{theorem}\label{thm:robust}
    Under conditions (1)-(5), and in the setting in which $m(y|x) = \int \phi_{\sigma_0}(y|\theta)f(\theta|x)\mu(d\theta)$ for some $\sigma_0\in\Sigma$, it is the case that $\log L_n/n \rightarrow \eta_0 - K(\sigma)$ almost surely for a constant $\eta_0$ and a continuous function $K(\sigma)$ such that $K(\sigma)>0$ if and only if $\sigma \neq \sigma_0$. Furthermore, this convergence is uniform, and the PRMLx estimator $\hat{\sigma}_n := \text{arg max}_{\sigma\in\Sigma}L_n(\sigma)$ converges almost surely to $\sigma_0$.
\end{theorem}
That is, so long as the set $\Sigma$ contains $\sigma_0$, we can rest assured knowing that the PRMLx estimator will be consistent. This result is analogous to that of Theorem 3 in \cite{martin2011semiparametric}, and establishes the PRMLx estimate as a straightforward means of tuning unmixed parameters that arises as a byproduct of the weight-localized predictive recursion algorithm.

While PRMLx is most naturally equipped to deal with hyperparameters of the kernel $\phi_\sigma$, they have in practice also been used to handle other nuisance parameters. For example, running weight-localized predictive recursion requires the input of a sequence of localization parameters $b_j$, which appears in the localization kernel. This value is of great importance, as it controls the degree to which data points near our choice of $x$ ``influence'' the predictive recursion algorithm. By augmenting the space $\Sigma$ with additional elements $b_j$ taking values on a closed set, one could in practice use the resulting estimate $\hat{b}_j$ as a principled ``guess'' at the best value of $b_j$. This is shown to have good empirical results in Section \ref{sec:sims}.
\section{Discussion}
PRx stands out among Bayesian density regression methods for its exceptional speed and reproducibility. Where competing procedures such as LSBP or SBART struggle with mixing and a reliance on initial values, the recursive scheme of PRx produces consistent results in a fraction of the time, while achieving stability through a simple permutation-averaging procedure. In comparison to this, we found that LSBP and SBART required significantly more runtime to process even moderately-sized data sets, and oftentimes resulted in noticeably different estimates between runs. As mentioned in Section \ref{sec:PRalgo}, the speed of PRx is a product of the \textit{predictive Bayes} interpretation of ordinary predictive recursion \citep{newton1998nonparametric, newton2002nonparametric}, which induces the sequence of predictive distributions that defines the algorithm. To the best of our knowledge, PRx is the first density regression procedure to exploit this framework, and thus benefits from it greatly in terms of speed.

It should be noted that conditional density estimation can also be framed from a frequentist perspective. The most popular example of this comes from \cite{de2003conditional}, who propose using the kernel conditional density estimate (KCDE):
\begin{equation}
\begin{split}
\hat{f}(y|x) = \frac{\sum_{i}K_{h_1}(y - y_i)K_{h_2}(\lVert x - x_i\rVert^2)}{\sum_i K_{h_2}(\lVert x - x_i\rVert^2)}
\end{split}
\end{equation}
where $K_h$ is some kernel with localization bandwidth $h$. Such procedures are fast and converge given a proper value for the bandwidth parameters $h_1$ and $h_2$ --- but in practice these quantities are not given. In fact, the crux of the literature on KCDE focuses on data-driven bandwidth selection; \cite{holmes2012fast} proposes a solution in the form of a cross-validation algorithm. PRx, on the other hand, comes with built-in bandwidth optimization in the form of PRMLx maximization, which is accessible as a byproduct of the PRx algorithm --- thus, there is no need to rely on a wholly separate procedure for estimating $b_j$. Beyond this, we believe that the semiparametric setup of PRx offers a level of interpretability and adaptability lacking in frequentist density regression procedures. The explicit mixture model underlying PRx allows the user to incorporate domain knowledge through the parametric function $\phi_\sigma(y|\theta)$, while KCDE forces a rigid, uninterpretable kernel over the $\mathcal{Y}$-space. Furthermore, a natural byproduct of using PRx that is not supported by KCDE is the recovery of a latent conditional mixing density, $f(\theta|x)$, which in many applied settings may be an object of interest. As seen in Sections \ref{sec:skew} and \ref{sec:testing}, PRx can also be adapted for a variety of inferential tasks, such as covariate-dependent multiple testing and Bayesian model comparison; in contrast, KCDE is designed explicitly as a means of density estimation. In lieu of these advantages, we believe that what PRx offers far exceeds the capabilities of frequentist density regression methods.

While the PRx algorithm itself is extremely fast, it should be noted that PRMLx maximization (Section \ref{sec:theory}) may suffer from a slower runtime in large-sample settings. In particular, computing the pseudo-likelihood function $\prod_{i=1}^n m_{i-1}(y_i|x_i)$ requires running PRx up to the $i-1$th iterate $n$ times, which scales as $O(n^2)$. When using iterative maximization algorithms such as L-BFGS-B---in which the PRMLx must be repeatedly computed---this can lead to slow runtimes. A simple way to remedy this issue is to run PRMLx maximization over a subset of the data. After obtaining estimates of the unmixed parameters and localization bandwidths using this subset, they can be plugged into the PRx algorithm and applied to all the data. More principled means of fast PRMLx maximization for large-sample data may be possible via nearest-neighbor type procedures, but for now we leave this avenue open for future research.

This work extends both the methodology and the theory behind predictive recursion to the covariate-dependent setting. The generality of kernel mixtures allows this procedure to adapt to a very broad class of densities, and the numerical results demonstrate that these estimates are comparable to those achieved by other Bayesian density regression methods. This, combined with a versatile model setup and fast runtime, makes PRx a practical and powerful tool for handling semiparametric mixture models.

\bibliographystyle{abbrvnat}
\bibliography{mybib}

\end{document}